\documentclass[aps,prl,twocolumn,showpacs,groupedaddress]{revtex4}

\usepackage{graphicx}
\usepackage{latexsym}

\usepackage{color}

\begin{document}

\title{Cis-Trans Dynamical Asymmetry in Driven Polymer Translocation }

\author{Takuya Saito}
\email[Electric mail:]{saito@fukui.kyoto-u.ac.jp}
\affiliation{Fukui Institute for Fundamental Chemistry, Kyoto University, Kyoto 606-8103, Japan}

\author{Takahiro Sakaue}
\email[Electric mail:]{sakaue@phys.kyushu-u.ac.jp}
\affiliation{Department of Physics, Kyushu University 33, Fukuoka 812-8581, Japan}

\def\Vec#1{\mbox{\boldmath $#1$}}
\def\degC{\kern-.2em\r{}\kern-.3em C}

\def\SimIneA{\hspace{0.3em}\raisebox{0.4ex}{$<$}\hspace{-0.75em}\raisebox{-.7ex}{$\sim$}\hspace{0.3em}} 

\def\SimIneB{\hspace{0.3em}\raisebox{0.4ex}{$>$}\hspace{-0.75em}\raisebox{-.7ex}{$\sim$}\hspace{0.3em}}

\date{\today}

\begin{abstract}
During polymer translocation driven by e.g. voltage drop across a nanopore, the segments in the cis-side is incessantly pulled into the pore, which are then pushed out of it into the trans-side. This pulling and pushing polymer segments are described in the continuum level by nonlinear transport processes known, respectively, as fast and slow diffusions. By matching solutions of both sides through the mass conservation across the pore, we provide a physical basis for the cis and trans dynamical asymmetry, a feature repeatedly reported in recent numerical simulations. We then predict how the total driving force is dynamically allocated between cis (pulling) and trans (pushing) sides, demonstrating that the trans-side event adds a finite-chain length effect to the dynamical scaling, which may become substantial for weak force and/or high pore friction cases.

\end{abstract}


\def\degC{\kern-.2em\r{}\kern-.3em C}

\newcommand{\gsim}{\hspace{0.3em}\raisebox{0.5ex}{$>$}\hspace{-0.75em}\raisebox{-.7ex}{$\sim$}\hspace{0.3em}} 
\newcommand{\lsim}{\hspace{0.3em}\raisebox{0.5ex}{$<$}\hspace{-0.75em}\raisebox{-.7ex}{$\sim$}\hspace{0.3em}} 

\maketitle

The dynamics of translocation, i.e., the polymer passage through a narrow pore, has been actively studied more than a decade~\cite{PNAS_Kasianowicz_1996,NanoLett_Storm_2005}. In addition to its relevance to cellular biological processes, i.e., biopolymer transports in cells, the phenomenon has found promising applications in genome sequencing and related technology as a nanopore sensor~\cite{Nature_Branton_2008,PNAS_Reiner_2010}. 
Being a unique mode of the molecular transport inherent in long flexible polymers, there have been numerous attempts to characterize the process and to uncover the underlying physics behind it~\cite{EPL_Luo_2009,EPL_Lehtola_2009,PRE_Bhattacharya_Binder_2010,JCP_Luo_2006,PRE_Saito_Sakaue_2012,PRL_Sung_1996,JCP_Muthukumar_1999,JPhys_Panja_2007,JCP_Haan_2012,PRE_Kantor_2004,PRE_Sakaue_2007,JPhys_Vocks_Panja_2008,EPJE_Saito_Sakaue_2011,JPCB_Rowghanian_Grosberg_2011,PRE_Dubbeldam_2012,PRE_Ikonen_Sung_2012,JStatMech_Panja_2010_01}. Such efforts have led to the consensus that the {\it tension} is a key physical quantity, which is created in the polymer by the spontaneous segment motions and/or the action of the driving force~\cite{PRE_Kantor_2004,PRE_Sakaue_2007,EPJE_Saito_Sakaue_2011,JPCB_Rowghanian_Grosberg_2011,PRE_Dubbeldam_2012,PRE_Ikonen_Sung_2012}.
Now we have a fairly good description of the translocation process, which may be categorized according to the magnitude $f$ of the bias as follows.
(i) Unbiased regime:  the tension imbalance across the pore arising from the segment exchange between cis and trans sides creates a long term power-law decaying memory~\cite{JPhys_Panja_2007,JCP_Haan_2012,JStatMech_Panja_2010_01}. This retards the process, leading to the anomalous (sub-) diffusion of the so-called translocation coordinate~\cite{PRL_Sung_1996,JCP_Muthukumar_1999}.
(ii) Weakly driven regime: with weak enough force, the polymer conformation is essentially in equilibrium, and the same physics as that in the unbiased regime applies. The linear response theory then leads to the anomalous drift of the translocation coordinate~\cite{JPhys_Vocks_Panja_2008,JStatMech_Panja_2010_01}.
(iii) Driven regime: when the rate of the driving operation is faster in comparison with the terminal time of the polymer, the genuine nonequilibrium dynamics shows up~\cite{PRE_Sakaue_2007,EPJE_Saito_Sakaue_2011,JPCB_Rowghanian_Grosberg_2011,PRE_Dubbeldam_2012,PRE_Ikonen_Sung_2012,PRE_Saito_Sakaue_2012}. The action of the driving force is transmitted along the chain in the cis-side, which induces the sequential conformational deformation. 
Here to make the connection to the regime (ii) clearer, it is convenient to introduce the length scale $\xi_f \simeq k_\mathrm{B}T/f$ and the corresponding time scale $\tau_f \simeq \tau_0 (\xi_f/a)^z$ \cite{PRE_Sakaue_2013} with $a$, $k_\mathrm{B}T$ being the segment size, thermal energy, respectively, and $z$ is a dynamical exponent. The shortest (segment scale) time scale is denoted as $\tau_0 \simeq \eta a^3/k_\mathrm{B}T$, where $\eta$ is the solution viscosity.
Initially, the process takes place locally in the close proximity to the pore, and the equilibrium treatment is valid up to $t = \tau_f$, during which the tension is transmitted up to $g_f \simeq (\xi_f/a)^{1/\nu}$-th monomer according to the mechanism in regime (ii). The subsequent larger scale process is governed by the driving force, thus, described by the nonequilibrium regime (iii), which eventually dominates the scaling limit. This consideration provides the border between the regimes (ii) and (iii) as $ g_f \simeq N_0 \Leftrightarrow f \simeq k_\mathrm{B}T/(aN_0^{\nu})$, the threshold force being rather weak for long chains, where $N_0$ is the polymerization index of the chain. For strong enough  force ($f \gtrsim k_\mathrm{B}T/a$), one only has the regime (iii) in the entire process.

\begin{figure}[t]
\begin{center}
\includegraphics[scale=0.70]{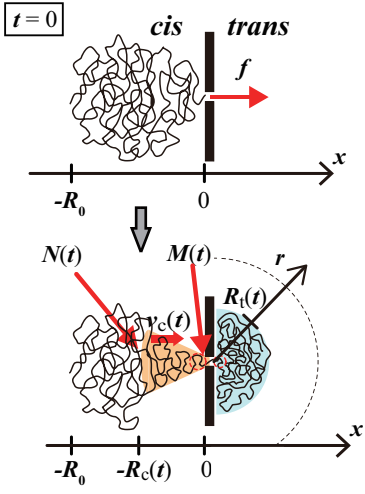}
      \caption{
	  (Color Online) Illustration of driven polymer translocation process.
	  Stretched and compressed domains are shaded in different colors.
	  }
\label{fig1}
\end{center}
\end{figure}

We remark here an important distinction between regimes (ii) and (iii). While in equilibrium regime (ii), one can treat the cis and trans sides on even ground, a qualitative difference should unclose as a characteristic feature in the nonequilibrium driven regime (iii). Here, compared to the sequential stretching of the cis-side chain owing to the propagating tension, the chain portion, which is pushed out of the pore faster than it relaxes in the trans-side, is compressed and resides dynamically in the crowded state.
While the proper description of the cis-side dynamics constitutes a basis for our current understanding of the dynamical scaling in the driven translocation~\cite{PRE_Sakaue_2007,EPJE_Saito_Sakaue_2011,JPCB_Rowghanian_Grosberg_2011,PRE_Dubbeldam_2012,PRE_Ikonen_Sung_2012,PRE_Saito_Sakaue_2012}, the trans-side event has been mostly overlooked so far.  
Although the occurrence of the dynamical crowding has repeatedly been addressed in recent literature e.g. 
through the observation of simulation snapshots~\cite{EPL_Luo_2009} or the numerical results~\cite{PRE_Bhattacharya_Binder_2010,PRE_Dubbeldam_2012}, it is poorly understood, and one is left with the question as to how the process in the trans-side can be described and  when it becomes important. Below, we attempt to provide the answer to it.

The driving force $f$ sucks the cis-side polymer segments into the pore, then pushes them into the trans-side.  
\begin{eqnarray}
f = f_\mathrm{c} + f_\mathrm{t} + f_\mathrm{pore}.
\end{eqnarray}
The polymer is stretched in the cis-side by the force $f_\mathrm{c}$, while it is dynamically compressed in the trans-side by $f_\mathrm{t}$, and the pore plays a role of the sink and the source in the cis- and trans-sides, respectively. There is also a frictional force $f_\mathrm{pore}$ associated with the pore.
To handle such a dynamically asymmetric situation on an equal footing, we introduce a length scale $\xi ({\vec r})$;  in the cis-side, this length is a tensile blob inversely proportional to the local chain tension, while it corresponds to the concentration blob in the trans-side, from which we can identify the thermodynamic force and the transport coefficient in the respective domains.
Below the length scale $\xi$, the chain conformation is essentially in equilibrium, i.e., letting $g({\vec r})$ be the segment number constituting the blob, $\xi ({\vec r}) \simeq a g ({\vec r})^{\nu}$~\cite{deGennesBook}.
In larger scale, the asymmetry shows up in such a way that the problem is essentially one-dimensional in the cis-side, while it is three-dimensional in the trans-side.
Keeping such a difference in the effective dimensionality $d$ in mind, let us introduce 
 the dimensionless segment density $\phi_d({\vec r}) \simeq a^d g({\vec r})/\xi({\vec r})^d \sim \xi({\vec r})^{p_{\nu, d}}$, where $d=1$ and $d=3$ for the cis- and trans-sides, respectively, and $p_{\nu,d} \equiv (1-d \nu)/\nu$.
 Given the free energy density $f_d({\vec r}) \simeq k_\mathrm{B}T/\xi({\vec r})^d$ and the transporting coefficient $L({\vec r}) \simeq  g({\vec r})/(\eta \xi({\vec r}))$ evaluated as a Stokes friction for each blob (non-draining case)~\cite{deGennesBook,EPL_Brochard_1995},  the constitutive relation for the segment flux ${\vec j}_d({\vec r})$ and the thermodynamic force can be standardized as
 \begin{eqnarray}
 {\vec j}_d({\vec r}) 
 &=& -L ({\vec r}) \phi_d({\vec r}){\vec \nabla}_d\left( \frac{\partial f_d}{\partial \phi_d}\right) 
 \nonumber \\
 &=& -D_0 \phi_d({\vec r})^{-p_z/p_{\nu,d}} {\vec \nabla}_d\phi_d({\vec r}),
 \label{flux}
 \end{eqnarray}
 where ${\vec \nabla}_d$ is a $d$-dimensional Nabla operator, and $D_0 \simeq k_\mathrm{B}T/\eta a$ is the segmental diffusion coefficient, and we introduce an exponent $p_z = z-2 \,(>0)$, which enables us to treat the effect of hydrodynamic interactions collectively, i.e., setting $z=3$ reduces to the above non-draining case, while $z= 2+\nu^{-1}$ represents the free-draining (Rouse) dynamics~\cite{deGennesBook,PRE_Sakaue_2007}. 
With the continuity equation of segments, this leads to the nonlinear diffusion equation of the type of porous media equation~\cite{PNAS_Barenblatt_2000}. Note that the exponent $p_{\nu,d} \equiv (1-d \nu)/\nu$ has different signs in the cis- and trans-sides~\cite{PRL_Sakaue_2009,PRE_Sakaue_2012,PRE_Rowghanian_Grosberg_2012,Macromolecules_Paturej_2012}. From eq.~(\ref{flux}), the case with plus (minus) sign indicates the enhanced~\cite{PRE_Sakaue_2012,PRE_Rowghanian_Grosberg_2012,Macromolecules_Paturej_2012} (suppressed~\cite{PRL_Sakaue_2009}) flux in less dense region. Such processes are called fast (slow) diffusions, which turns to be crucial in the following.
 
It is also useful to introduce the velocity potential
 \begin{eqnarray}
 \psi({\vec r}) 
 \equiv \frac{p_{\nu,d}}{p_z} D_0\phi_d({\vec r})^{-p_z/p_{\nu,d}} 
 \simeq \frac{p_{\nu,d}}{p_z} D_0 \left(\frac{\xi({\vec r})}{a}\right)^{-p_z},
 \end{eqnarray}
the gradient of which leads to the segment velocity ${\vec \nabla}_d \psi({\vec r}) = {\vec j}_d({\vec r})/\phi_d({\vec r}) ={\vec v}_d({\vec r})$, hence its naming.

\begin{figure}[t]
\begin{center}
\includegraphics[scale=0.70]{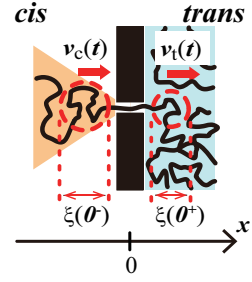}
      \caption{
	  (Color Online) Enlarged illustration in the vicinity of pore.
	  Dashed circles represent closest blobs to the pore on the cis-/trans-sides. 
	  }
\label{fig2}
\end{center}
\end{figure}

Take $x$ coordinate perpendicular to a thin wall with a small hole, which is located at the origin (Fig.~1). We set the cis-side in $x<0$ region, where the polymer is initially placed taking an equilibrium conformation. At $t=0$, the translocation process begins, when one of the chain ends finds the pore. The segments are sequentially labeled from the first arriving end. At time $t$, $M(t)$-th segment is located at the pore, which is traditionally called a translocation coordinate in literature~\cite{PRL_Sung_1996,JCP_Muthukumar_1999}, while $N(t)$-th segment is at the rear end of the moving domain, which represents the dynamics of the tension propagation in the cis-side (see below and Fig.~1).

{\it Cis-side dynamics ($d=1$): fast diffusion}---
Our dynamical equation in $d=1$ dimension describes the stretching process~\cite{PRE_Sakaue_2012,PRE_Rowghanian_Grosberg_2012,Macromolecules_Paturej_2012}. The essential nonequilibrium feature here can be captured by what we call the two-phase picture, in which the entire cis-side polymer is divided into the moving domain with a characteristic velocity $v_\mathrm{c}$ and the yet quiescent domain (Fig.~1).  Integration of the velocity potential in the moving domain range, we have
\begin{eqnarray}
\int_{-R_\mathrm{c}}^{0^{-}} \mathrm{d}x \ v_1(x) &=& \psi(0^{-}) - \psi(-R_\mathrm{c})  \nonumber \\
&\simeq& D_0 \left[ \left( \frac{\xi(0^{-})}{a}\right)^{-p_z} - \left( \frac{\xi(-R_c)}{a}\right)^{-p_z} \right],
\nonumber\\
\end{eqnarray}
where $x=-R_\mathrm{c}$ is the location of the tension propagation front (see Fig.~1), and the positive numerical constant $p_{\nu,1}/p_z$ of order unity is absorbed in $D_0$.
Given the nonuniform stretched conformation $\xi(0^-) \ll \xi(-R_\mathrm{c})$, one can obtain the force-velocity relation for the cis-side polymer
\begin{eqnarray}
v_\mathrm{c}(t) R_\mathrm{c}(t) \simeq D_0 \left( \frac{f_\mathrm{c}(t) a}{k_\mathrm{B}T}\right)^{p_z},
\label{v_c}
\end{eqnarray}
where the boundary conditions $v(0^-) \simeq v_\mathrm{c}(t)$ (see ref.~\cite{JPCB_Rowghanian_Grosberg_2011} and \cite{EPJE_Saito_Sakaue_2011} arXiv) and $\xi(0^-) = \xi_{f_\mathrm{c}} \simeq k_\mathrm{B}T/f_\mathrm{c}(t)$ are used (see Fig.~2).
In addition, from the definition of the tension front, we have the relation
\begin{eqnarray}
R_\mathrm{c}(t) \simeq a N(t)^{\nu},
\label{init}
\end{eqnarray}
which traces back to the initial equilibrium conformation at $t=0$.

{\it Trans-side dynamics ($d=3$): slow diffusion}---
The trans-side decompression process can be described by our dynamical equation in $d=3$ dimension~\cite{PRL_Sakaue_2009}. Taking the spherical symmetry into account, the integration of the velocity potential leads to
\begin{eqnarray}
\int_{0^{+}}^{R_\mathrm{t}} \mathrm{d}r \ v_3(r) &=&  \psi(R_\mathrm{t}) -\psi(0^{+})  \nonumber \\
&\simeq& D_0 \left[ \left( \frac{\xi(0^{+})}{a}\right)^{-p_z} - \left( \frac{\xi(R_\mathrm{t})}{a}\right)^{-p_z} \right],
\end{eqnarray}
where $r=R_{\rm t}$ is the radial position of the decompressed front of the trans-side chain (see Fig.~1), and the positive numerical constant $-p_{\nu,3}/p_z$ of order unity is absorbed in $D_0$.
To proceed, we assume that the velocity field rapidly decreases away from the pore so that the above integral can be dominated by the pore vicinity. 
We then find 
\begin{eqnarray}
v_\mathrm{t}(t) a \simeq D_0 \left( \frac{f_\mathrm{t}(t) a}{k_\mathrm{B}T}\right)^{p_z+1},
\label{v_t}
\end{eqnarray}
where the boundary conditions $v(0^+) = v_\mathrm{t}(t)$ and $\xi(0^+) \simeq k_\mathrm{B}T/f_\mathrm{t}(t)$ are used.
Equation~(\ref{v_t}) states that the force-velocity relation in the trans-side is regularized locally as the condition for the blob closest to the pore (see Fig.~2). 
This may be justified by the physical observation that the injection rate is controlled by the local condition in the pore vicinity, but not aware of the position of the decompression front $R_\mathrm{t}(t)$. 
The contrast to eq.~(\ref{v_c}), which includes the global information $R_c(t)$, reflects a qualitative difference between the pulling and pushing operations.
Indeed, this first blob is pushed by $f_{\rm t}$, which is balanced by the drag force against it, i.e., $f_\mathrm{t}\sim \xi_\mathrm{t}^{p_z}v_\mathrm{t}$.
Putting $f_\mathrm{t} \sim \xi_\mathrm{t}^{-1}$ into it leads to eq.~(\ref{v_t}).

\begin{figure}[t]
\begin{center}
\includegraphics[scale=0.45]{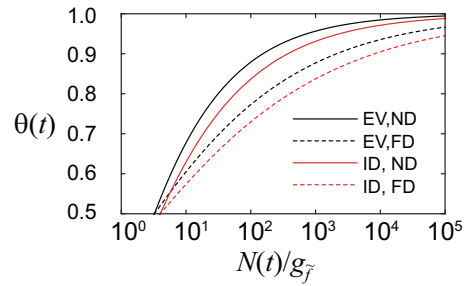}
      \caption{
	  (Color online) Single logarithmic plot of fraction of the cis-side force $\theta (t)=A^{-1}(N(t)/g_{\tilde{f}})$ in eq.~(\ref{decoupled_A}).
	  The Flory and dynamical exponents are adopted as follows; ideal chain (ID) $\nu=1/2$, excluded volume chain (EV) $\nu=0.588$, free-draining (FD) $z=2+\nu^{-1}$ and non-draining (ND) $z=3$.
	  }
\label{fig3}
\end{center}
\end{figure}

{\it Mass conservation through pore}---
We now patch the above described cis- and trans-sides dynamics by requiring the conservation of mass across the pore. The rate of segments sucked from the cis-side into pore is $\mathrm{d}M(t)/\mathrm{d}t = [\phi_1(0^-)/a] v_1(0^-)$, which is equal to the rate of segments pushed into the trans-side $[\phi_3(0^+)/a^3] v_3(0^+) \xi(0^+)^2$. Using the boundary conditions at the pore, it can be rewritten as
\begin{eqnarray}
a \frac{\mathrm{d}M(t)}{\mathrm{d}t} \simeq  \left( \frac{f_\mathrm{c}(t) a}{k_\mathrm{B}T}\right)^{-p_{\nu, 1}} v_\mathrm{c}(t) \simeq \left( \frac{f_\mathrm{t}(t) a}{k_\mathrm{B}T}\right)^{-p_{\nu, 1}} v_\mathrm{t}(t).
\label{flux_pore}
\end{eqnarray}
Eliminating $v_\mathrm{c}$ and $v_\mathrm{t}$ using eqs.~(\ref{v_c}) and~(\ref{v_t}), we obtain the expression of the tension front $R_\mathrm{c}(t)$, hence $N(t)$ via Eq.~(\ref{init}), in terms of the force allocation.  
\begin{eqnarray}
N(t) \simeq g_{f_\mathrm{c}(t)} \left( \frac{f_\mathrm{c}(t)}{f_\mathrm{t}(t)}\right)^{(p_z-p_{\nu,1}+1)/\nu},
\label{fc_ft}
\end{eqnarray}
where, as already defined, $g_{f_\mathrm{c}(t)} \simeq (f_\mathrm{c}(t) a /k_\mathrm{B}T)^{-1/\nu}$ is the number of segments in the blob immediate vicinity of the pore in the cis-side. To get a clear-cut time dependence of the fraction of the cis-side force $\theta(t) = f_\mathrm{c}(t)/{\tilde f}$, where ${\tilde f} = f-f_\mathrm{pore} = f_\mathrm{c}(t) + f_\mathrm{t}(t)$, one can rewrite eq.~(\ref{fc_ft}) as
\begin{eqnarray}
N(t) /g_{{\tilde f}} \simeq A(\theta(t))
\end{eqnarray}
with $g_{{\tilde f}} \simeq ({\tilde f} a /k_\mathrm{B}T)^{-1/\nu}$ and the function
\begin{eqnarray}
A(\theta) \equiv 
\theta^{(p_z-p_{\nu,1})/\nu}(1-\theta)^{-(p_z-p_{\nu,1}+1)/\nu},
\label{decoupled_A}
\end{eqnarray}
which is monotone increasing given the domain $\theta \in (0,1)$ (in practice $\theta \ge 0.5$, however. See below.)
In Fig.~3, we plot $\theta(t)$ as a function of $N(t)$. At $t \simeq \tau_f$, which serves as the initial condition of the nonequilibrium driven regime (iii), the chain in the cis-side forms an initial tensed blob. This is connected to our remark in the introduction that, until this moment, cis- and trans-sides can be treated on even ground, thus, $f_{\rm c} = f_{\rm t}$. Along with translocation process advanced, the tension propagates in cis-side, and $f_\mathrm{c}$ monotonically increases and eventually dominates in magnitude over $f_\mathrm{t}$. Since the normalization factor of $N(t)$ is $g_{{\tilde f}}$, the growth of $\theta(t)$ is unconcerned with the total chain length $N_0$, thereby, the trans-side effect is irrelevant in the dynamical scaling scenario with the identification $f_\mathrm{c} = {\tilde f}$ in the long chain limit~\cite{FN_Ikonen}. 
It also predicts the smaller the driving force and/or the larger the pore friction, the larger the finite-chain length effect, since both of these make $g_{\tilde{f}}$ larger.
The translocation time in the asymptotic limit can be derived as follows.
In this limit, the tension propagation dynamics is described from eqs.~(\ref{v_c}),~(\ref{init}) and the first equality in eq.~(\ref{flux_pore}) 
\begin{eqnarray}
\frac{\mathrm{d}M(t)}{\mathrm{d}t} \simeq \tau_0^{-1} \left( \frac{{\tilde f}a}{k_\mathrm{B}T}\right)^{p_z-p_{\nu, 1}} N(t)^{-\nu}.
\end{eqnarray}
This equation can be closed by an additional relation 
$N(t) \simeq M(t)$ valid in the driven regime (iii) (see ref.~\cite{JPCB_Rowghanian_Grosberg_2011} and \cite{EPJE_Saito_Sakaue_2011} arXiv). Then, the tension propagation time $\tau$ is identified as $N(\tau) = N_0$;
\begin{eqnarray}
\tau \simeq \tau_0 \left( \frac{{\tilde f}a}{k_\mathrm{B}T}\right)^{p_{\nu, 1}-p_z} N_0^{1+\nu},
\label{tau}
\end{eqnarray}
which coincides in leading order to the translocation time.
As is evident from Fig. 3, the approach to the asymptote $f_\mathrm{c}  \simeq {\tilde f}$ is rather slow, which may cause the subtlety and difficulty to measure the force exponent in eq.~(\ref{tau}).
Finally, note $g_{{\tilde f}} \rightarrow N_0$ in the limit of the weak force ${\tilde f} \rightarrow k_\mathrm{B}T/(aN_0^{\nu})$, thus, we have a crossover to the weakly driven regime (ii) at this force, where the cis- and trans-sides can be treated on even ground during the whole translocation process.

In summary, we have provided a lucid description on the cis and trans dynamical asymmetry in the driven translocation. The imbalance is further promoted under higher driving force ($f>k_\mathrm{B}T/a$), which leads to the almost full stretching in the cis-side. 
In such situations, we only need to modify exponents in the transport
equation in cis-side as $p_z=1$ and $p_{\nu}=0$~\cite{PRE_Sakaue_2012}. 
For our purpose, this requires replacing the counterparts in eq.~(\ref{v_c}) and the first equation of eqs.~(\ref{flux_pore}).
We can then show a similar result with $A(\theta)=\theta^{1/\nu} (1-\theta)^{-(p_z-p_{\nu,1}+1)/\nu}$ and $g_{\tilde{f}}=(\tilde{f}a/k_\mathrm{B}T)^{-(p_z-p_{\nu,1})/\nu}$.
Note that $g_{\tilde{f}}$ no longer has the meaning of the initial tensed blob in this strong force regime; $g_{\tilde{f}} \simeq 1$ at $\tilde{f} \simeq k_\mathrm{B}T/a$ and getting smaller with the force, which indicates the faster approach to the asymptote $\theta=1$.
A basic equation (eq.~(\ref{flux})) can also be applied to the polymer transport in different space dimensions with appropriate exponents $\nu$ and $z$. Therefore, the same asymmetric features would be verified in the translocation process, for instance, conducted in the slit geometry.

\section*{Acknowledgements}

This work was supported by the JSPS Core-to-Core Program
``Non-equilibrium dynamics of soft matter and information" and JSPS KAKENHI Grant Number 24340100.


\begin{thebibliography}{21}

\bibitem{PNAS_Kasianowicz_1996}
	{
	J.~J.~Kasianowicz, E.~Brandin, D.~Branton and D.~W. Deamer,
  Proc. Natl. Acad. Sci. U.S.A.
  {\bf 93}, 13770 (1996).
	}

\bibitem{NanoLett_Storm_2005}
 {A.~J. Storm, C. Storm, J. Chen, H. Zandbergen, J.-F. Joanny and C. Dekker,
   Nano Lett. 
   {\bf 5}, 1193 (2005).}

\bibitem{Nature_Branton_2008}
	{D. Branton et. al., Nature Biotechnology, {\bf 26}, 1146 (2008).}


\bibitem{PNAS_Reiner_2010}
	{J.~E. Reiner, J.~J. Kasianowicz, B.~J. Nablo and J.~W.~F. Robertson, 
	Proc. Natl. Acad. Sci. {\bf 107}, 12080 (2010).}   
   
   
  
\bibitem{EPL_Luo_2009}
	{ K.~Luo, T.~Ala-Nissila, S.~-C.~Ying and R.~Metzler,
  Europhys. Lett. {\bf 88}, 68006 (2009).}

  
\bibitem{EPL_Lehtola_2009}
	{ V.~V.~Lehtola, R.~P.~Linna and K.~Kanski,
  Europhys. Lett. {\bf 85}, 58006 (2009).}

\bibitem{JCP_Luo_2006}
	{ I. Huopaniemi, K.~Luo, T.~Ala-Nissila and S.~-C.~Ying,
  J. Chem. Phys. {\bf 125}, 124901 (2006).}  
  
  
  
\bibitem{PRE_Bhattacharya_Binder_2010}
	{A.~Bhattacharya and K.~Binder,
  Phys. Rev. E {\bf 81}, 041804 (2010).}






\bibitem{PRL_Sung_1996}
	{ W. Sung and P.~J. Park,
  Phys. Rev. Lett. {\bf 77}, 783 (1996).}

\bibitem{JCP_Muthukumar_1999}
	{M. Muthukumar,
  J. Chem. Phys. {\bf 111}, 10371 (1999).}

\bibitem{JPhys_Panja_2007}
	{ D. Panja, G.~T. Barkema, R.~C. Ball, 
	J. Phys.: Condens. matter {\bf 19}, 432202 (2007).}

\bibitem{JCP_Haan_2012}
	{H.~W de Haan and G.~W. Slater, 
	J. Chem. Phys. {\bf 136}, 154903 (2012).}

\bibitem{JStatMech_Panja_2010_01}
  {D. Panja,
	J. Stat. Mech. L02001 (2010); ibid L06011 (2010).}   	
  
\bibitem{JPhys_Vocks_Panja_2008}
{ H.~Vocks, D.~Panja, G.~T.~Barkema and R.~C.~Ball,
  J. Phys.: Condens. Matter {\bf 20}, 095224 (2008).}

  
\bibitem{PRE_Kantor_2004}
	{ Y. Kantor and M. Kardar,
  Phys. Rev. E {\bf 69}, 021806 (2004).}  


\bibitem{PRE_Sakaue_2007}
	{ T.~Sakaue,
  Phys. Rev. E {\bf 76}, 021803 (2007); ibid {\bf 81}, 041808 (2010).}	    
  		


\bibitem{EPJE_Saito_Sakaue_2011}
	{T. Saito and T. Sakaue,
	Eur. Phys. J. E {\bf 34}, 135 (2011); ibid {\bf 35}, 125 (2012). See also arXiv:1205.3861
	}


	
	
	
\bibitem{JPCB_Rowghanian_Grosberg_2011}
	{ P. Rowghanian and A. Y. Grosberg,
  J. Phys. Chem. B {\bf 115}, 14127-14135 (2011).}

\bibitem{PRE_Dubbeldam_2012}
	{ J.~L.~A. Dubbeldam, V.~G. Rostiashvili, A. Milchev and T.~A. Vilgis
  Phys. Rev. E {\bf 85} 041801 (2012).}  
  
  
\bibitem{PRE_Ikonen_Sung_2012}
	{ T. Ikonen, A. Bhattacharya, T.~Ala-Nissila and W. Sung,
  Phys. Rev. E {\bf 85} 051803 (2012).}  
  

\bibitem{PRE_Saito_Sakaue_2012}
	{ T. Saito and T. Sakaue,
  Phys. Rev. E {\bf 85}, 061803 (2012).}    


  
\bibitem{PRE_Sakaue_2013}
  {T. Sakaue,
	Phys. Rev. E, {\bf 87}, 040601(R) (2013)}   
		
\bibitem{deGennesBook}
	{ P.-G.~de~Gennes,
  {\it Scaling Concepts in Polymer Physics}
  (Cornell University Press, Ithaca, 1979).}

\bibitem{EPL_Brochard_1995}
	{F.~Brochard-Wyart,
  Europhys. Lett. {\bf 30}, 387 (1995).}  
  

\bibitem{PNAS_Barenblatt_2000} 
   { G. I. Barenblatt, M. Bertsch, A. E. Chertock, and V. M. Protokishin, 
   Proc. Natl. Acad. Sci. USA {\bf 97}, 9844 (2000).}
		
  
\bibitem{PRE_Sakaue_2012}
	{ T. Sakaue, T. Saito and H. Wada,
  Phys. Rev. E {\bf 86}, 011804 (2012).}    


  
\bibitem{PRE_Rowghanian_Grosberg_2012}
	{ P. Rowghanian and A. Y. Grosberg,
  Phys. Rev. E {\bf 86}, 011803 (2012).}   
  
\bibitem{Macromolecules_Paturej_2012}
  {J. Paturej, A. Milchev, V. G. Rostiashvili, and T. A. Vilgis,
		Macromolecules {\bf 45}, 4371 (2012).}



	
		
\bibitem{PRL_Sakaue_2009}
	{ T. Sakaue and N. Yoshinaga,
  Phys. Rev. Lett. {\bf 102}, 148302 (2009).}    
		



  
\bibitem{FN_Ikonen}
	{
	The same remark was made in recent work~\cite{PRE_Ikonen_Sung_2012}, where the trans-side effect is treated assuming the near-equilibrium statistics.
	}  

  




  

\end{thebibliography}
\end{document}